\documentclass[a4paper,11pt]{article}
\pdfoutput=1 

\usepackage{jcappub} 

\usepackage[T1]{fontenc} 
\usepackage{color}
\usepackage{hyperref}

\hypersetup{
    colorlinks=true,
    linkcolor=red,
    citecolor=blue,
} 

\newcommand{\be}{\begin{equation}}
\newcommand{\ee}{\end{equation}}
\newcommand{\bs}{\begin{split}} 
\newcommand{\bea}{\begin{eqnarray}}
\newcommand{\eea}{\end{eqnarray}} 
\newcommand{\mpl}{M^2_{\rm pl}} 
\newcommand{\ms}{M^2_{\star}} 
\newcommand{\ml}{M_\Lambda}

\newcommand{\eps}{\epsilon} 
\newcommand{\lamt}{\lambda^3}

\title{The Well-Tempered Cosmological Constant: Fugue in B$^\flat$} 

\author[a]{Stephen Appleby,}
\affiliation[a]{Asia Pacific Center for Theoretical Physics, Pohang, 37673, Korea}
\author[b,c]{Eric V.~Linder}
\affiliation[b]{Berkeley Center for Cosmological Physics \& Berkeley Lab, 
University of California, Berkeley, CA 94720, USA}  
\affiliation[c]{Energetic Cosmos Laboratory, Nazarbayev University, 
Nur-Sultan, Kazakhstan 010000}

\abstract{
Zero point fluctuations of quantum fields should generate a large cosmological 
constant energy density in any spacetime. How then can we have 
anything other than de Sitter space without fine tuning? 
Well tempering -- dynamical cancellation of the cosmological constant using degeneracy within the field equations -- can replace a large cosmological constant with a much lower energy state. Here we give 
an explicit mechanism to obtain a Minkowski solution, 
replacing the cosmological constant with zero, and testing 
its attractor nature and persistence through a vacuum 
phase transition. 
We derive the general conditions that Horndeski 
scalar-tensor gravity must possess, and evolve in a fugue of functions, to deliver nothing and make the universe 
be flat. 
}

\begin{document}
\maketitle
\flushbottom


\section{Introduction} 

Vacuum energy is an essential element of quantum theory. Zero point 
fluctuations exist, and have an energy density associated with them 
even if the vacuum expectation value of the field is otherwise zero. 
This issue is problematic in cosmology, where the vacuum energy, i.e.\ 
cosmological constant, should 
gravitate and hence affect the curvature of spacetime and the evolution 
of the universe \cite{Weinberg:1988cp} (for recent reviews see \cite{Carroll:2000fy,Martin:2012bt,Nobbenhuis:2004wn,Padilla:2015aaa}).

One of the few ways of dealing with this satisfactorily 
in a cosmological scenario is well tempering \cite{temper1,1812.05480,temper2}. 
(See
\cite{Dolgov:1983,ArkaniHamed:2002fu,Kaloper:2013zca,Kaloper:2014fca,Kaloper:2015jra,Brax:2019fgj,Sobral-Blanco:2020rdu,Lombriser:2019jia,Amariti:2019vfv,Evnin:2018zeo} for some different approaches to the problem.) Here the 
evolution equations of the background spacetime and of the field are 
degenerate, providing a potential dynamical cancellation of the large 
cosmological constant, generically of order the Planck energy, or 
possibly some other high energy  phase transition. Well tempering has 
been applied to cancel the large cosmological constant and replace it 
with a late time, much lower energy de Sitter state -- in order to 
explain the observed late time cosmological acceleration. 

However suppose we want to apply it outside a cosmological context, 
and not replace the large cosmological constant with another constant 
energy density, but rather with nothing, to have a Minkowski spacetime, 
the familiar stage for (flat spacetime) quantum field theory. Indeed, 
the first papers on self tuning \cite{self1,self2}, the predecessor to well 
tempering, worked with a Minkowski background  (also see 
\cite{self3,Gubitosi:2011sg,fab5,fab5b,Appleby:2015ysa}). 

We explore the ability of well tempering to deliver nothing, to make 
the universe be flat despite a large vacuum energy. In 
Sec.~\ref{sec:method} we lay out well tempering in a Minkowski context, 
using the general Horndeski formulation of scalar-tensor gravitation. 
We will see that interactions between the different terms in the Lagrangian 
throughout the evolution, a fugue of functions, is an essential element. 
Section~\ref{sec:g3} presents solutions of the field equations for 
models with no coupling between the scalar field and the Ricci scalar curvature (cubic Horndeski) while 
Sec.~\ref{sec:g4} includes explicit coupling to curvature. The soundness 
of the theory, in terms of freedom from ghosts and Laplace 
instability, and demonstration of cancellation through a 
phase transition, is treated in Sec.~\ref{sec:sound} and Sec.~\ref{sec:phase} respectively. 
We conclude in 
Sec.~\ref{sec:concl}.

\section{Minkowski Well Tempering} \label{sec:method} 
Our starting point is the action \cite{Horndeski:1974wa,Nicolis:2008in,Deffayet:2009wt,Deffayet:2011gz} 
\begin{equation} 
S = \int d^{4}x\,\sqrt{-g}  \left[ {1 \over 2}(M_{\rm pl}^{2} + M \phi)  R + K(X) - \lambda^{3}\phi - G_{3}(X)\Box\phi - \Lambda \right] \,, 
\end{equation} 
where $\phi$ is a scalar field with kinetic term $X=-(1/2)g^{\mu\nu}\nabla_\mu\phi\nabla_\nu\phi$ and $\Lambda$ is the cosmological constant. The general form of the action has been extensively used for dark energy and cosmic inflation \cite{Deffayet:2010qz,DeFelice:2010nf,Kobayashi:2011nu}, but is studied within a different context in this work.

Since we don't want to solve one 
quantum problem only to have it overtaken by another -- quantum 
loop corrections -- we employ only shift symmetric terms. Therefore the Horndeski 
Lagrangian terms have $G_4=(\mpl+M\phi)/2$ (so a shift 
in $\phi$ is absorbed into $\mpl$), 
$G_3=G_3(X)$, and the kinetic/potential term is $K(X)-\lamt\phi$. The field equations are given by \cite{Bernardo:2020ehy} 
\begin{eqnarray} 
\label{eq:eins} & & {M_{\rm pl}^{2} \over 2} G_{\mu\nu} - {M \over 2} \left(\phi_{\mu\nu} - g_{\mu\nu}\Box\phi \right) - {1 \over 2} g_{\mu\nu} K + {1 \over 2} g_{\mu\nu} \lambda^{3}\phi - {1 \over 2} \phi_{\mu}\phi_{\nu}K_{X} + \\ 
\nonumber & & \qquad {1 \over 2} \left[ \phi_{\mu} \phi_{\nu} \Box\phi - \phi_{\mu}\phi^{\lambda}\phi_{\nu\lambda} - \phi_{\nu}\phi^{\lambda}\phi_{\mu\lambda}  + g_{\mu\nu} \phi^{\lambda}\phi^{\beta} \phi_{\lambda\phi}\right] G_{3 X}  + {1 \over 2} g_{\mu\nu} \Lambda = 0 \\
\label{eq:sfe} & & K_{X} \Box \phi - K_{XX} \phi^{\alpha}\phi^{\beta}\phi_{\alpha\beta} - \lambda^{3} + {M \over 2} R + \\ 
\nonumber  & & \qquad G_{3 X} \left[ \phi_{\alpha\beta}\phi^{\alpha\beta} + \phi^{\alpha}\Box \phi_{\alpha} - \phi^{\alpha}\nabla_{\alpha}\Box\phi  - (\Box\phi)^{2} \right]  + G_{3 XX}\phi^{\beta}\phi_{\alpha\beta}\left[ \phi^{\alpha}\Box\phi - \phi^{\lambda}\phi_{\lambda}{}^{\alpha} \right] = 0 \,, 
\end{eqnarray} 
where subscripts on the field $\phi$ indicate covariant derivatives $\phi_{\mu\nu} \equiv \nabla_{\mu}\nabla_{\nu}\phi$ and 
a subscript $X$ denotes a derivative 
with respect to $X$. 

Well tempering is a mechanism by which exact vacuum solutions are constructed for which the metric is independent of the value of the cosmological constant $\Lambda$. In this paper we are looking for solutions to the Einstein equations that are identically Minkowski space. If we fix the metric using an ansatz $g_{\mu\nu} = \eta_{\mu\nu}$, where  $\eta_{\mu\nu} = {\rm diag}(-1,1,1,1)$ is the Minkowski metric, then typically the coupled scalar field and Einstein equations will be over-constrained. To evade this, we require some form of degeneracy within the dynamical equations. This can be achieved either by enforcing that the scalar field equation ($\ref{eq:sfe}$) is identically satisfied by the metric ansatz \cite{self1,self2}, or alternatively by demanding that the scalar field and metric dynamical equations are equivalent  \cite{temper1}. Each of these scenarios can be realized with an appropriate choice of scalar field terms and derivative couplings in the Lagrangian. 

In \cite{temper1}, the authors searched for well tempered de Sitter solutions for which the expansion rate of the spacetime is independent of $\Lambda$. In this work we seek Minkowski solutions (hence we are not trying to do 
observational cosmology). To this end, 
we take $g_{\mu\nu} = \eta_{\mu\nu}$ and $\phi = \phi(t)$. The equations greatly simplify, and we obtain the $(\mu,\nu)=(0,0)$, $(i,j)$ Einstein and scalar field equations as 
 \begin{eqnarray} 
 & & K  - K_{X}  \dot{\phi}^{2} - \lambda^{3}\phi  =  \Lambda  \label{eq:ham}\\ 
 & & \left( -  K + G_{3 X}  \dot{\phi}^{2}\ddot{\phi} + \lambda^{3}\phi - M  \ddot{\phi}  \right) \eta_{ij} = -\eta_{ij}\Lambda \label{eq:hdot}\\
 & & - K_{X} \ddot{\phi} - K_{XX} \dot{\phi}^{2} \ddot{\phi} -\lambda^{3}  = 0\,. \label{eq:field}
\end{eqnarray} 
Adding the first and second equations gives the more convenient 
\be 
\ddot\phi\,(M-2XG_{3X})+2XK_X=0\,. \label{eq:hdot2}
\ee 

As expected, the equations are over-constrained as the field $\phi$ must satisfy both Eqs.~($\ref{eq:field}$) and ($\ref{eq:hdot2}$). However, by selecting the functions $G_{3}$ and $K$ appropriately, such that Eqs.~($\ref{eq:field}$) and ($\ref{eq:hdot2}$) are identical `on-shell'\footnote{Adopting the terminology of \cite{self1}, `on-shell' means that the metric is exactly vacuum $g_{\mu\nu} = \eta_{\mu\nu}$.}, an exact solution to the field equations can be derived in which the metric is exactly flat spacetime, and the field $\phi(t)$ evolves and dynamically eliminates the effect of $\Lambda$. The field $\phi(t)$ does not relax to a constant vacuum expectation value, and no parameters in the action require fine tuning. This is the crux of well tempering. 

In the following sections we provide two cases of this approach, models that possess no explicit coupling between $\phi$ and $R$ and that do include coupling, with 
explicit examples.

\section{Fugue in B Flat Minor: $G_3+K$} \label{sec:g3} 

First we consider only the kinetic/potential term 
$K(X)-\lamt\phi$ and cubic term $G_3(X)$ in the Lagrangian. In the absence of any explicit coupling between $\phi$ and $R$, 
i.e.\ $M=0$, 
Eqs.~($\ref{eq:field}$) and ($\ref{eq:hdot2}$) are 
 \begin{eqnarray}  
 \label{eq:deg1} & &  - G_{3 X}   \dot{\phi}^{2}\ddot{\phi} + K_{X} \dot{\phi}^{2} = 0 \\
 \label{eq:deg2} & & - K_{X} \ddot{\phi} - K_{XX} \dot{\phi}^{2} \ddot{\phi} - \lambda^{3} = 0\,.  
 \end{eqnarray} 
To obtain a consistent solution, these two equations must be equivalent, which can be arranged with an appropriate choice of $K$ and $G_{3}$. This is the {\it degeneracy condition\/}, and will be written explicitly below. 

There are two immediate consequences to 
this requirement. 
For these equations to be equivalent we need the coefficients of the $\ddot\phi$ terms in both equations 
to be nonzero. Therefore we require both $K \neq 0$ and $G_3 \neq 0$ -- it 
will be the interplay between the two, the fugue, that enables 
well tempering. 
Second, the tadpole $\lamt\phi$ is essential, as without it the scalar field equation admits a constant field solution $\ddot{\phi} = 0$.  The role of the tadpole was similarly critical in \cite{temper1}, in a different context.

To establish the degeneracy condition, we write each equation ($\ref{eq:deg1}),  (\ref{eq:deg2}$) 
as $\ddot\phi=\,$RHS and equate the two right hand sides, yielding the degeneracy condition 
\be 
G_{3X}=-\frac{1}{\lamt}\,K_X(K_X+2XK_{XX})\,.  \label{eq:g3fromk}
\ee 

Any model that satisfies this condition will admit an exact Minkowski solution with a dynamical scalar field $\phi(t)$ canceling the vacuum energy. Given some $K(X)$ we derive $G_3$ from Eq.~(\ref{eq:g3fromk}). The scalar field equation can then be solved as 
\be 
\int dX\,\frac{K_X+2XK_{XX}}{X^{1/2}}=-\lamt\sqrt{2}\int dt\,. 
\ee  
The simplest example is $K=\eps X$, where $\eps$ is a constant 
(not equal to zero; it can be set to 1 for a canonical kinetic 
term). This choice results in 
\bea 
G_3&=&-\frac{\eps^2}{\lamt}X  \label{eq:g3nog4can}\\ 
\phi&=&-\frac{\lamt}{2\eps}t^2+c_1t+c_0\,, \label{eq:phican} 
\eea  
where $c_0$, $c_1$ are constants of  integration. 
It is a striking result that a simple cubic Galileon action, containing a canonical kinetic term, tadpole, and $(\partial \phi)^{2} \Box\phi$ can successfully screen an arbitrary vacuum energy. In this model, the scalar field evolves indefinitely according to Eq.~($\ref{eq:phican}$). 

More generally, models with non-standard kinetic terms $K_X=\eps X^n$ (note $\eps$ is dimensionful for $n \neq 0$), have the same well tempering property if they are accompanied by a corresponding $G_{3}$ function  
\bea 
G_3&=&-\frac{\eps^2}{\lamt}\,X^{2n+1}\\ 
\dot\phi&=&-\left|-\frac{\lamt 2^n}{\eps}\,t+c_1\right|^{1/(1+2n)}\,. \label{eq:phin}
\eea 
Note that $\ddot\phi$ has the same sign as 
$-\lamt/[\eps(1+2n)]$. 

For this class of models, the Hamiltonian constraint Eq.~(\ref{eq:ham}) reduces to 
\be 
\lamt d_0+\Lambda=0\, ,  \label{eq:hamn} 
\ee 
where $d_{0}$ is related to the integration constants $c_{0}, c_{1}$. Hence $\Lambda$ is canceled by arbitrary constants that are fixed by solving the Hamiltonian constraint as an initial condition for $\phi$, $\dot{\phi}$. No fine tuning of $\phi$ or $\dot{\phi}$ is required, any choice of initial data is acceptable, subject only to the Hamiltonian constraint being satisfied. By solving Eq.~(\ref{eq:ham}) initially, $\Lambda$ will remain canceled throughout the subsequent evolution (we emphasize there is no fine tuning: Eq.~\ref{eq:hamn} can be thought of as a relation between $c_0$ and $c_1$, which enter the time dependent function $\phi(t)$; the cancellation is an attractor if we start off shell; and the cancellation occurs even if $\Lambda$ undergoes a phase transition, as we show 
in Sec.~\ref{sec:phase}). 
The field $\phi$ does not relax to a constant vacuum expectation value but continues dynamically to evolve on the `vacuum' by virtue of the degeneracy condition within the field equations\footnote{Note that all solutions for the dynamical field $\phi(t)$ are invariant under the transformation $t \to t + t_{0}$ for constant $t_{0}$, which respects the symmetry of the underlying spacetime. However, when we impose the Hamiltonian constraint, the arbitrary integration constants $c_{1}, c_{0}$ are partially fixed in terms of physical mass scales $\Lambda$ and $\lambda$.}.

As an alternative method of constructing well tempered models, we can use 
Eq.~(\ref{eq:g3fromk}) to choose $G_3(X)$ and solve for $K$ via 
\bea  
\frac{d(XK_X^2)}{dX}&=&-\lamt G_{3X}\\ 
K_X&=&X^{-1/2}\,(c-\lamt G_3)^{1/2}\,. 
\eea   
For $G_{3}=-(\eps^2/\lamt)X$, with $c=0$, we return to the simple case $K=\eps X$. Keeping $c=0$, with $G_3=-(\eps^2/\lamt)X^n$, we have 
\bea 
K_X&=&\eps X^{(n-1)/2}\\ 
\dot\phi&=&\left[-\frac{\lamt}{\eps} 2^{(n-1)/2}\,t+c_1\right]^{1/n}\,. 
\eea 

Showing the existence of a Minkowski vacuum solution is not the end of the story, because this background might be unstable either dynamically or to metric and scalar field perturbations. Therefore, we test the dynamical stability of the simplest model presented in this section, initially to determine if the Minkowski vacuum state is an attractor. To this end, we allow the spacetime metric to be dynamical. Specifically we insert a flat 
Friedmann-Lema{\^\i}tre-Robertson-Walker (FLRW) metric into the covariant equations, perturb the metric away from its flat space limit and study the evolution of the expansion rate $H = \dot{a}/a$. Taking an FLRW metric   
\begin{equation} 
ds^{2} = - dt^{2} + a^{2}(t) \delta_{ij}dx^{i}dx^{j} \,, 
\end{equation} 
and well tempered model $K = \epsilon X$, $G_{3} =- \epsilon^{2} X/\lambda^{3}$, the field equations read 
\bea 
3M_{\rm pl}^{2} H^2 &=& M_{\Lambda}^{4} + {\epsilon \over 2} \dot{\phi}^{2} + \lambda^{3} \phi - 3{\epsilon^{2}  \over \lambda^{3}} H \dot{\phi}^{3}
\label{eq:fried}\\ 
-2M_{\rm pl}^{2}\dot H &=& 
{\epsilon^{2} \dot{\phi}^{2} \over \lambda^{3}} \ddot\phi - 3 {\epsilon^{2} \over \lambda^{3}}H\dot{\phi}^{3} + \epsilon \dot{\phi}^{2} \label{eq:fulldh}\\ 
0&=&\left( \epsilon  - {6\epsilon^{2} H \dot{\phi} \over \lambda^{3}} \right) \ddot\phi
+3\epsilon H\dot\phi + \lambda^{3} - {3 \epsilon^{2} \over \lambda^{3}} \dot{\phi}^{2} (\dot H+3H^2)\,, \label{eq:fullddphi} 
\eea 
where we have defined $M_{\Lambda}^{4} \equiv \Lambda$. To solve this system numerically, we fix $\epsilon = 1$, $M_{\Lambda} = \lambda = 1$, $M_{\rm pl} = 10^{2}M_{\Lambda}$. With this choice, we are effectively scaling all dimensionful units in the system by $M_{\Lambda}$, and fixing $\lambda = M_{\Lambda}$. This choice is arbitrary; in general $\lambda$ is a free mass scale. 

For late time cosmology, we are interested in the regime in which there exists a hierarchy of scales $M_{\rm pl} \gg M_{\Lambda} \gg H$. We therefore randomly select pairs of random initial conditions over the range $10^{-2} < \dot{\phi}_{i}/M_{\Lambda}^{2} < 1$ and $10^{-4}< H_{i}/M_{\Lambda} < 10^{-2}$, and solve the Friedmann equation to obtain $\phi_{i}/M_{\Lambda}$ on the initial time slice. We then evolve Eqs.~(\ref{eq:fulldh})--(\ref{eq:fullddphi}) until $H/M_{\Lambda} < 10^{-5}$. We confirm that the Friedmann equation is solved on each timestep.

\begin{figure}
  \centering 
    \includegraphics[width=0.48\textwidth]{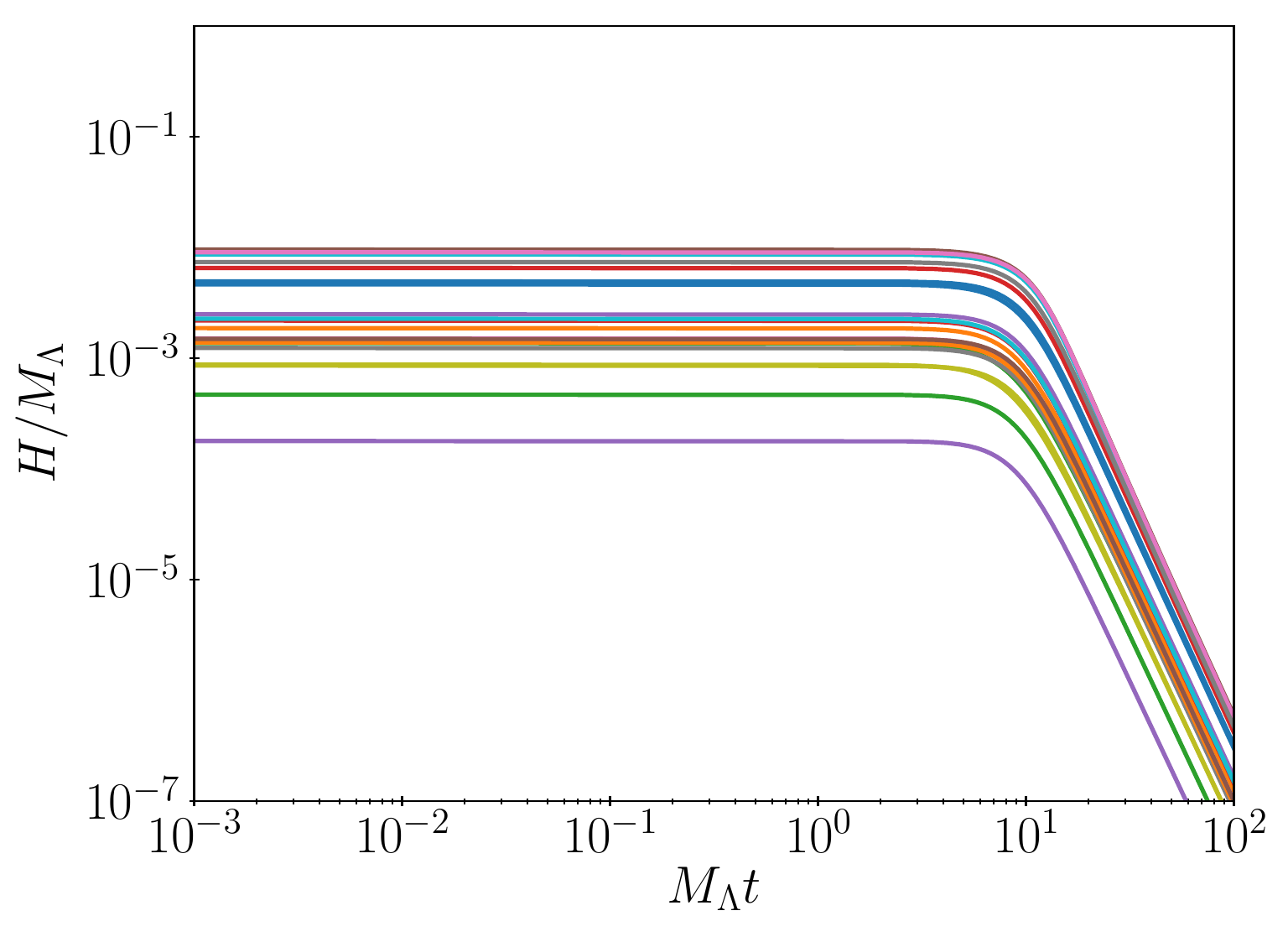}\\  
  \includegraphics[width=0.48\textwidth]{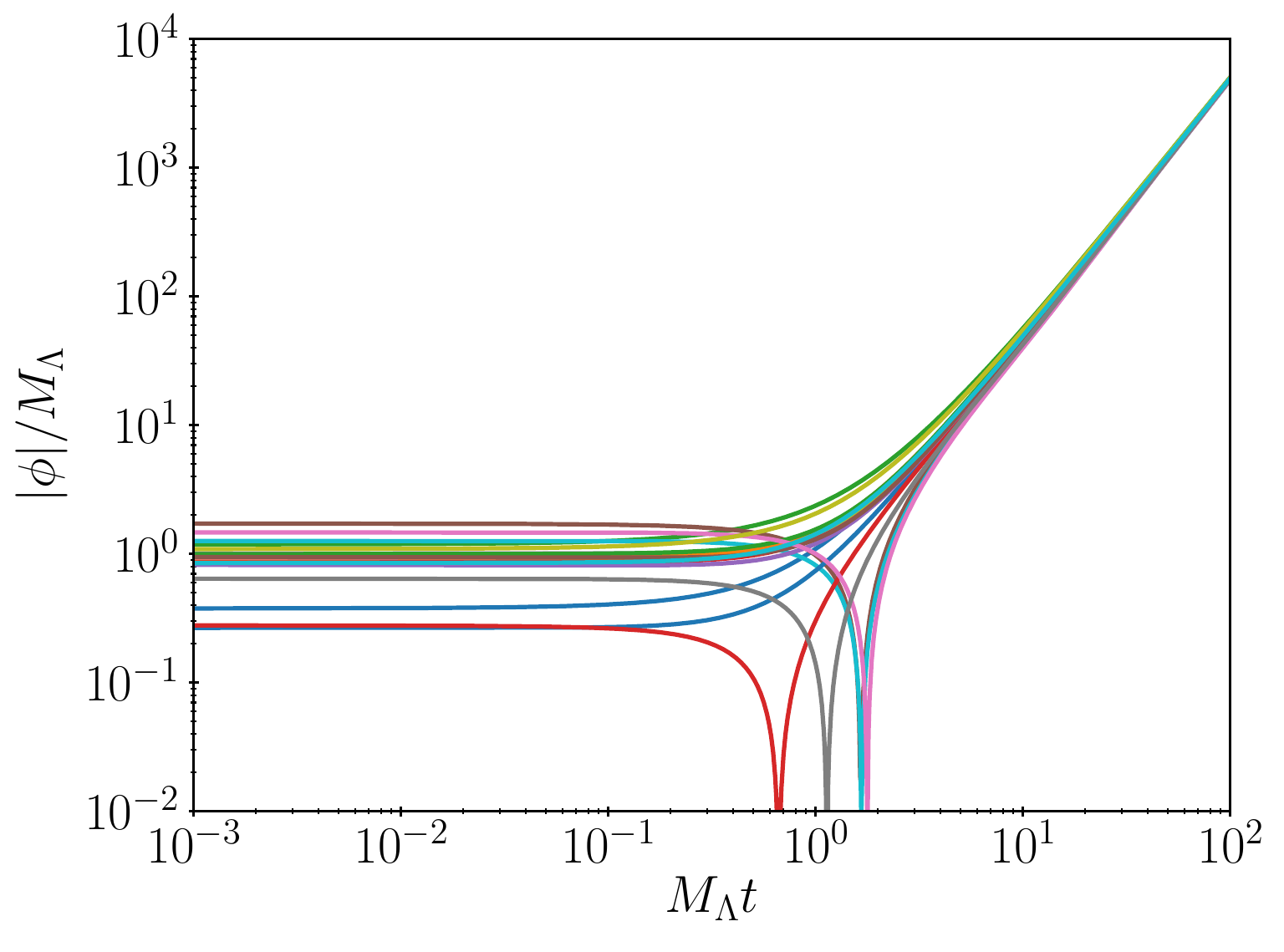}  
    \includegraphics[width=0.48\textwidth]{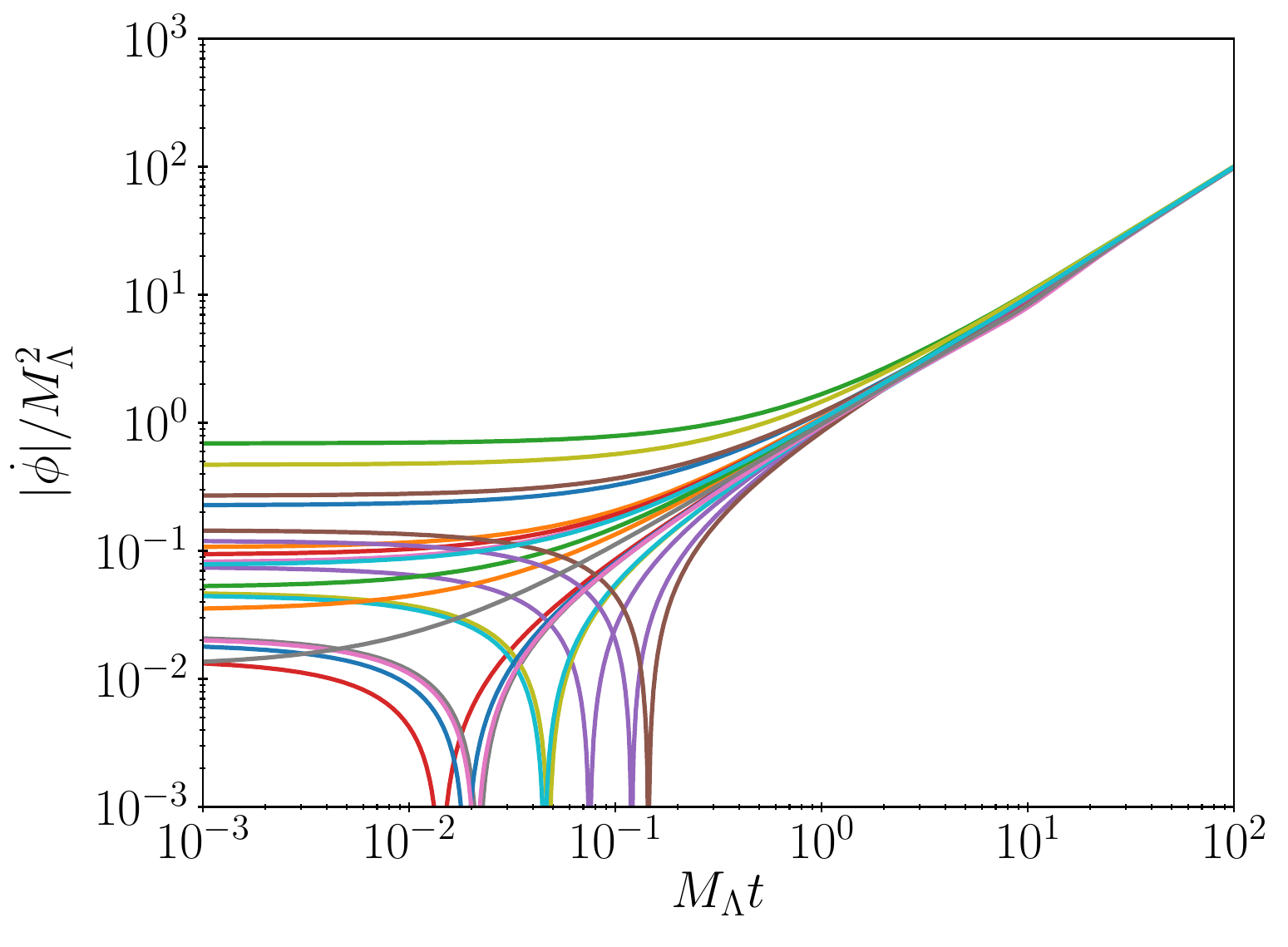} 
  \caption{
  Evolution for the well tempered $G_3+K$ case 
  (``B$^\flat$ minor'') with $K=X$. 
[Top panel] The evolution of $H$ as a function of $M_{\Lambda}t$ for a set of random initial conditions. The Hubble parameter loiters around its initial condition before evolving to the well tempered vacuum state $H \to 0$. 
  [Bottom panels] The corresponding evolution of the absolute value of the field $\phi$ (left) and $\dot\phi$ (right) as a function of $M_{\Lambda}t$, with random initial conditions for $\dot\phi$ (then $H_i$ and $\dot\phi_i$ determine $\phi_i$). Regardless of the initial conditions the field follows a consistent track in the parameter space, towards larger (negative) values.  The cusps in $\phi$ and $\dot{\phi}$ are zero crossings of the field. 
  } 
  \label{fig:1}
\end{figure}

Our results are presented in Fig.~\ref{fig:1}. Both the field $\phi$ and expansion rate $H$ loiter around the random initial condition, before approaching the degenerate Minkowski vacuum solution $H \to 0$ and $\phi \to -\lambda^{3}t^{2}/2\epsilon$. 
The transition occurs when the field reaches the attractor solution, at 
approximately $\ml t\sim  |\dot\phi_i/\ml|(\ml^3/\lamt)$. Once the attractor 
solution for $\dot\phi$ is reached, $\phi$ is moved toward its attractor solution around $\ml t\sim \ml^3/\lamt$, 
and then $H$ begins to decline toward zero as $t^4$. 
The well tempered solution is an attractor solution of the dynamical system. This will also be important in Sec.~\ref{sec:phase}. We discuss the stability of the model to inhomogeneous perturbations in Sec.~\ref{sec:sound}.

\section{Fugue in B Flat Major: $G_4+G_3+K$} \label{sec:g4} 

We now include an explicit coupling to curvature, between $\phi$ and the 
Ricci scalar, of the form $G_{4}(\phi) = (M_{\rm pl}^{2} + M\phi)/2$. The field dynamics is described by  
Eqs.~(\ref{eq:field})--(\ref{eq:hdot2}) and we can write Eq.~($\ref{eq:hdot2}$) as 
\be 
\ddot\phi=\frac{-2XK_X}{M-2XG_{3X}}\,. \label{eq:phiall} 
\ee 

\noindent The degeneracy  condition for the existence of a well-tempered vacuum solution becomes 
\begin{equation} 
\label{eq:min1} {2XK_{X} \over M - 2 X G_{3X}} = {\lambda^{3} \over K_{X} + 2X K_{XX}} \,. 
\end{equation} 

\noindent Once again the tadpole $\lambda^{3}$ is essential for the mechanism. Note that neither the Hamiltonian constraint Eq.~(\ref{eq:ham}) nor 
the field evolution equation (\ref{eq:field}) alters its 
form with the presence of $G_4$. Multiplying through by 
$\dot\phi$, we can write the field equation as 
\be 
K_X+2XK_{XX}=-\lamt\,\frac{d\phi}{dt}\frac{dt}{dX}\,, 
\ee 
which can be written effectively as  
\be 
-\frac{d(K-2XK_X-\lamt\phi)}{dX}=0\,, 
\ee 
i.e.\  
\be 
K-2XK_X-\lamt\phi={\rm const}\,. 
\ee

\noindent Conversely, one can start with this expression and take 
the time derivative to get Eq.~(\ref{eq:field}).  
Now looking at the 
Hamiltonian constraint Eq.~(\ref{eq:ham}), we see 
this is exactly what is needed to cancel $\Lambda$. 
Thus, unlike the general FLRW case 
where the field equation has terms with $H$ and $\dot H$, in the 
Minkowski case a well tempered model that satisfies the field equation 
automatically cancels a large cosmological constant. 

Finally, using the degeneracy condition we obtain 
\be 
G_{3X}=\frac{M}{2X}-\frac{1}{\lamt}\,K_X(K_X+2XK_{XX})\,. \label{eq:degall} 
\ee 
Thus we select some function $K(X)$, 
determine $G_{3}$ from Eq.~($\ref{eq:degall}$), and know that such a model can temper a large cosmological 
constant $\Lambda$ to Minkowski. 
(However, we cannot select
$K_X\sim X^{-1/2}$, as this yields no $\ddot\phi$ term in the scalar field equation.) 

Each choice of $K$ will have its own solution for 
$\phi(t)$, and this will be independent of $M$ at the  
background level. 
For example, for $K_X=\eps X^n$ then 
\be 
\dot\phi=\left(-\frac{\lamt \,2^n}{\eps}\,t+c_1\right)^{1/(1+2n)}\,. \label{eq:phikxn} 
\ee 
This is the same as  Eq.~(\ref{eq:phin}), but the presence of $G_4$ changes the well tempering expression for $G_3$. For $K=\eps X$ we now have 
\be  
G_3=-\frac{\eps^2}{\lamt}\,X+\frac{M}{2}\,\ln X\,, 
\ee 
modified from Eq.~(\ref{eq:g3nog4can}). 

Note there also exists a solution with $G_3=0$ but non-zero $M$, i.e.\ $G_4+K$, though it is not pretty: 
\bea 
K_X&=&X^{-1/2}\sqrt{\frac{M\lamt}{2}\ln X+k}\\ 
\dot\phi&=&e^{\frac{\ln 2}{2}-\frac{k}{M\lamt}}\,\exp\left\{\left(-\frac{\lamt}{\sqrt{2}}t+c\right)^2/(M\lamt)\right\}\,.  
\eea 

Following the analysis of the previous section, we again numerically evolve the dynamics of the simplest model considered in this section, with $K = \epsilon X$, $G_{3} = -\epsilon^{2} X/ \lambda^{3} +(M/2)\ln X$, and $G_{4} = (M_{\rm pl}^{2} + M \phi)/2$. The introduction of $M$ does not alter the background solution, but could change the dynamics when off the Minkowski vacuum state. Inserting an FLRW metric, the equations read 
\bea 
3(M_{\rm pl}^{2} + M\phi) H^2 &=& M_{\Lambda}^{4} + {\epsilon \over 2} \dot{\phi}^{2} + \lambda^{3} \phi - 3{\epsilon^{2}  \over \lambda^{3}} H \dot{\phi}^{3}
\label{eq:fried_wm}\\ 
-2(M_{\rm pl}^{2} + M\phi) \dot H &=& 
{\epsilon^{2} \dot{\phi}^{2} \over \lambda^{3}} \ddot\phi - 3 {\epsilon^{2} \over \lambda^{3}}H\dot{\phi}^{3} + \epsilon \dot{\phi}^{2} + 2 M H \dot{\phi}  \label{eq:fulldh_wm}\\ 
0&=&\left( \epsilon  - {6\epsilon^{2} H \dot{\phi} \over \lambda^{3}} \right) \ddot\phi
+3\epsilon H\dot\phi + \lambda^{3} - {3 \epsilon^{2} \over \lambda^{3}} \dot{\phi}^{2} (\dot H+3H^2) + 3 M H^{2}\,.  \label{eq:fullddphi_wm} 
\eea 
We repeat our analysis from Section~\ref{sec:g3}, selecting the same parameters $\lambda = 1 = M_{\Lambda}$, $M_{\rm pl} = 10^{2}M_{\Lambda}$ and $\epsilon = 1$. We additionally select three different values of $M=-M_{\Lambda}, 0, M_{\Lambda}$, and evolve the dynamical system assuming identical initial conditions for $H_{i}$, $\dot{\phi}_{i}$. The initial field value 
$\phi_{i}$ is fixed by the Friedmann equation, and varies for each choice of $M$.

\begin{figure}
  \centering 
  \includegraphics[width=0.48\textwidth]{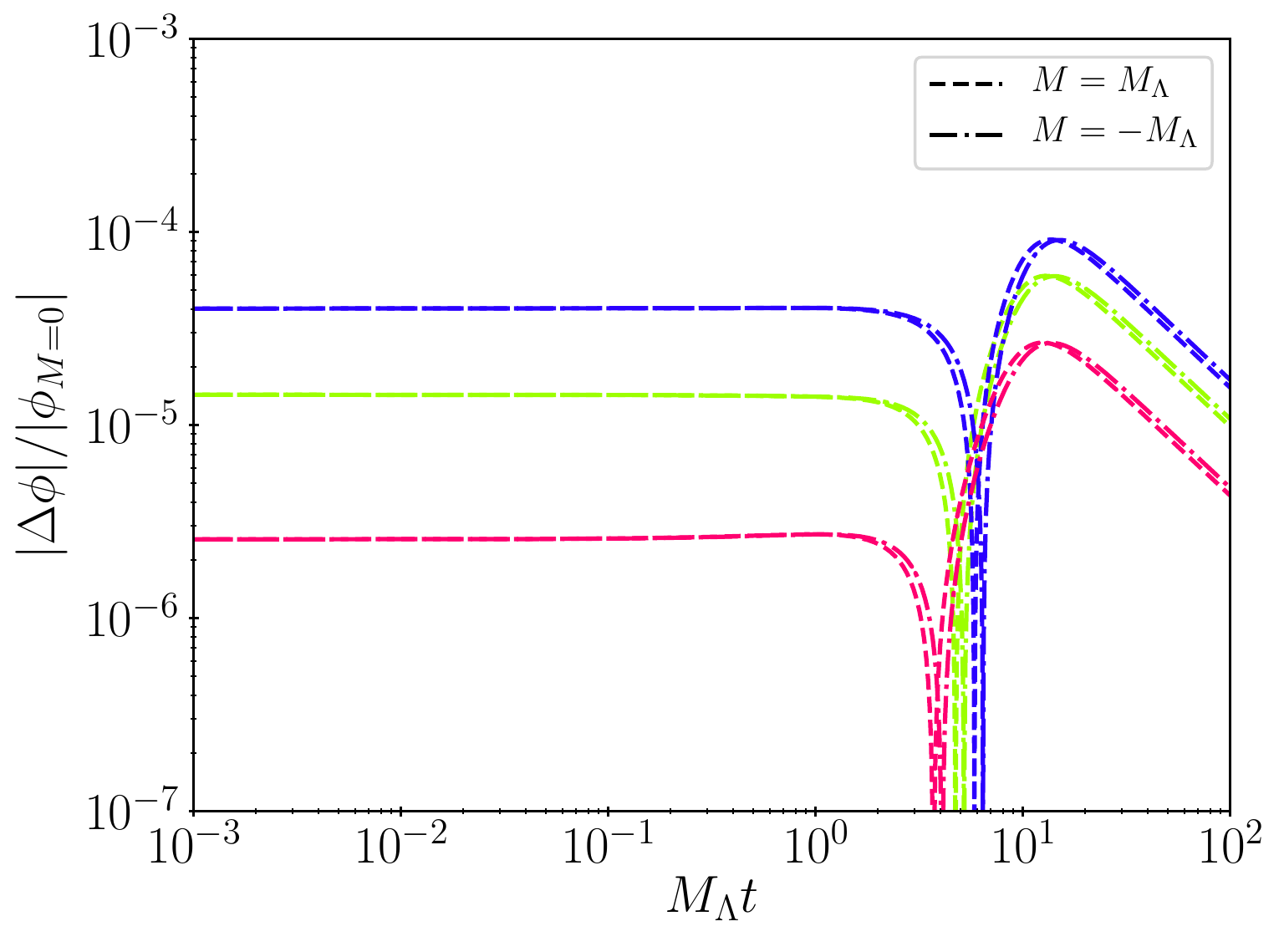}  
  \includegraphics[width=0.48\textwidth]{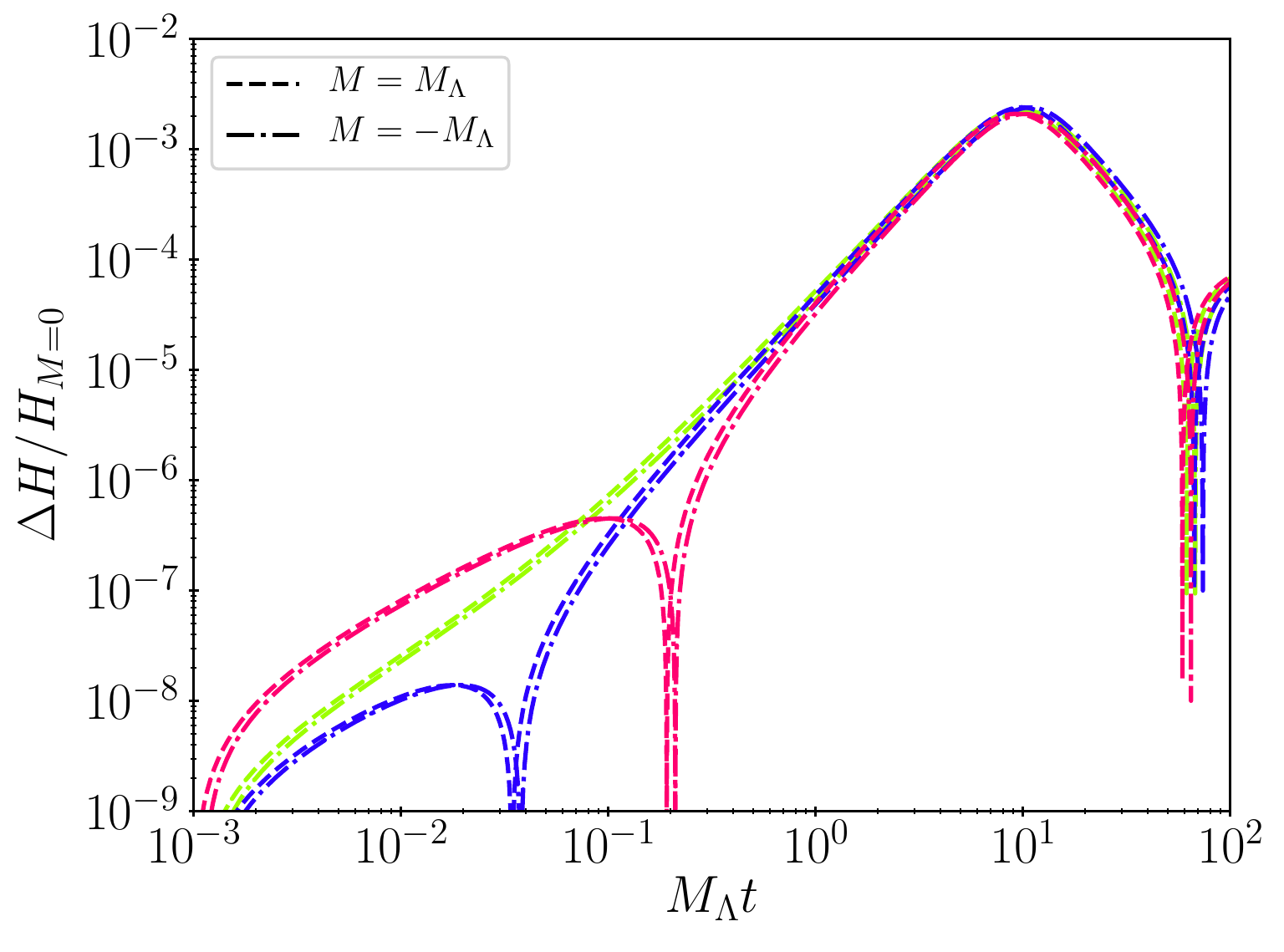}\\
  \caption{
  Evolution for the well tempered $G_4+G_3+K$ case
  (``B$^\flat$ major'') with $K=X$. 
  [Left panel] The fractional difference in field value  $\phi$ for the cases $M=0$ 
  and $M=M_{\Lambda}$ (dashed) and $M=-M_{\Lambda}$ (dot-dashed), as a function of $M_{\Lambda}t$ for three sets of random initial conditions.  [Right panel] The corresponding fractional deviation in $H$ as a function of $M_{\Lambda}t$. There is no significant dependence of the background dynamics on $M$, for reasonable parameter choices of $M_{\rm pl}$, $M$, and $M_{\Lambda}$.  
  }
  \label{fig:2}
\end{figure}

We present results for three randomly selected values of $H_{i}$, $\dot{\phi}_{i}$ in Fig.~\ref{fig:2}, relative to the uncoupled 
$M=0$ case. 
Specifically, we show $\Delta \phi / |\phi_{M=0}| = |\phi_{M = \pm M_{\Lambda}} - \phi_{M=0}|/|\phi_{M=0}|$ and  $\Delta H / H_{M=0} = |H_{M = \pm M_{\Lambda}} - H_{M=0}|/H_{M=0}$. The dashed/dot-dashed lines correspond to $M=M_{\Lambda}$ and $M = - M_{\Lambda}$ respectively, but nearly overlap. Each colored line represents a random initial condition. We observe no significant dependence of the evolution of the expansion rate on $M$; 
recall we already knew there was no impact on the field dynamics in the Minkowski limit. This arises because on the initial time slice we 
take $\mpl\gg M\phi$, i.e.\ the field starts subdominant to the Planck scale. Then at large $M_{\Lambda}t$, although 
the field grows and can become larger than $M_{\rm pl}$, this occurs on approach to the Minkowski vacuum solution, so any contribution to the dynamics from $M$ is suppressed by factors of $H, \dot{H} \ll 1$, as 
seen in the figure.

\section{Soundness} \label{sec:sound} 

Next we check that the well tempered vacuum solutions are classically stable, by 
evaluating the no ghost and Laplace (gradient) stability conditions. For 
freedom from ghosts in the shift symmetric Horndeski theory, one 
requires \cite{1107.3917,Bellini:2014fua}
\be 
\ms\left\{2\ms X(K_X+2XK_{XX})+12Xg^2-12MXg+3M^2X\right\}\ge0\,,  
\ee 
where we write $\ms=\mpl+M\phi$ and 
$g=XG_{3X}$. Normally $\ms>0$ but 
it is possible for it to be negative (recall we're 
dealing with Minkowski spacetime, not our universe),  
e.g.\ for large negative $\phi$. Below we assume 
$\ms>0$ unless otherwise stated, but it is 
straightforward to redo for the $\ms<0$ case. 
Putting in 
the well tempering condition Eq.~(\ref{eq:degall}) we find 
\be 
2X(K_X+2XK_{XX})\left[\ms+\frac{6}{\lambda^6}X^2K_X^2(K_X+2XK_{XX})\right]\ge 0\,. 
\quad [{\rm no\ ghost}]
\ee 
This holds, irrespective of $M$, as long as $K_X+2X K_{XX} \ge 0$, 
which we can arrange even if 
$K_X=\eps X^{n<-1/2}$ by taking $\eps<0$ (still giving 
$K>0$ for $n<-1$).  

The Laplace stability condition $c_s^2\ge0$ takes the form 
\cite{1107.3917} 
\be 
\ms\left\{\ms\ddot\phi\left[2g+4Xg_X-M\right]+X(M-2g)(3M+2g)\right\}\ge 0\,. 
\ee 
Substituting in for $\ddot\phi$ from Eq.~(\ref{eq:phiall}) 
and $g$ from the well tempering 
condition Eq.~(\ref{eq:degall}) yields 
\bea 
2\ms XK_X&+&\frac{4\ms X[XK_X(K_X+2XK_{XX})]_X}{(K_X+2XK_{XX})} 
+\frac{8M X^2K_X(K_X+2XK_{XX})}{\lamt}\notag\\
&-&\frac{4X^3K_X^2(K_X+2XK_{XX})^2}{\lambda^6}\ge0\,. 
\qquad [{\rm Laplace\ stability}]
\eea 
Thus $K$ must satisfy this condition to give Laplace stability. If this condition is violated, then perturbations will grow and ultimately render the Minkowski vacuum an unsuitable background.  

For the well tempered models, this condition often imposes limitations on 
the field excursion, as large excursions can  
render them asymptotically Laplace unstable. 
For example, for the model in 
Fig.~\ref{fig:1} the background spacetime is classically (Laplace) unstable to scalar perturbations at times $t\gtrsim (M_{\rm pl}/\lambda^{3})^{1/2}$. These particular vacuum states are not stable for arbitrarily large field excursions (beyond the Planck scale), but certain models below can be stable. 

For the case $K_X=\eps X^n$, we obtain 
\be 
\eps(3+4n)\ms  X^{1+n}+\frac{4M\eps^2(1+2n)}{\lamt}X^{2(1+n)} 
-\frac{2\eps^4(1+2n)^2}{\lambda^6}\,X^{3+4n}\ge 0\,. 
\ee 
If $n>-1/2$ then $X$ grows according to Eq.~(\ref{eq:phikxn}), 
and the last term ultimately dominates, violating the stability criterion. 
For $n<-1/2$, $X$ remains large if $\dot\phi$ has a pole, 
i.e.\ $-\lamt/\eps$ and $c_1$ are of opposite signs. Then 
the middle term will dominate for $-1<n<-1/2$. This then 
requires $\lamt<0$ for stability. For $n<-1$, the first term  
dominates and we require $\eps<0$ for stability. Finally, 
if $-\lamt/\eps$ and $c_1$ are both positive, then for 
$n<-1/2$ we have $X$ becoming small. For $n<-2/3$, the last 
term dominates and gives instability. For $-2/3<n<-1/2$ the 
first term dominates and gives stability if $\eps>0$. 
However, the no ghost condition requires $\eps<0$ so 
the power law $K_X$ nonpole class is not viable (and a pole ends evolution at a 
finite time). 
We can however verify that the well tempered $G_4+K$ case, an example of non-power law $K_X$, is both 
ghost free and Laplace stable for $K_X>0$.

 \section{Phase Transition}
\label{sec:phase}
 
Finally, we test that the well tempering models considered here can screen the vacuum energy even after a phase transition  in its value. To perform this check, we focus on the simplest model $K = \epsilon X$, $G_{3} = - \epsilon^{2}X/\lambda^{3}$, and coupling $M = 0$. 

The dynamical system in the presence of a phase transition can be described by the following equations 
 \bea 
3M_{\rm pl}^{2}H^{2} &=& \Lambda(t) + {\epsilon \over 2} \dot{\phi}^{2} + \lambda^{3} \phi - 3{\epsilon^{2}  \over \lambda^{3}} H \dot{\phi}^{3} \label{eq:pta1}\\ 
0&=&\left( \epsilon  - {6\epsilon^{2} H \dot{\phi} \over \lambda^{3}} \right) \ddot\phi
+3\epsilon H\dot\phi + \lambda^{3} - {3 \epsilon^{2} \over \lambda^{3}} \dot{\phi}^{2} (\dot H+3H^2)\,, \label{eq:pta2} 
\eea  
where $\Lambda$ is now explicitly time dependent. We assume that $\Lambda$ can be described with a step function \cite{self1}
 \begin{equation} 
 \Lambda = M_{\Lambda}^{4} + \Delta^{4} \theta(t - t_{c}) \,, 
 \end{equation} 
where $\theta(t - t_{c})$ is the Heaviside function and $t_{c}$ is the time of the transition. Such a discontinuity in the energy density will generically introduce divergences in $\ddot{\phi}$ and $\dot{H}$, however these divergences are not physical because we expect any realistic transition to occur over a finite timescale. 
For computational simplicity and clarity we use piece-wise continuous functions. 
 
We start the system at $t < t_{c}$ and assume that the fields are initially identically on-shell, with solution $H=0$, $\phi = -\lambda^{3}t^{2}/(2\epsilon) + c_{1} t + c_{0}$. The constants $c_1$, $c_0$ are related via 
\begin{equation} 
{c_{1}^{2} \epsilon \over 2} + c_{0}\lambda^{3} + M_{\Lambda}^{4} = 0 \,. 
\end{equation} 
This relation is the Hamiltonian constraint. At the phase transition $t = t_{c}$, $\phi$ is continuous (otherwise $\dot\phi$ diverges) but $H$ and $\dot{\phi}$ are not. It follows that $\ddot{\phi}$ and $\dot{H}$ diverge at this time. To obtain an analytic solution at this point, we rewrite the time derivative of Eq.~(\ref{eq:pta1}), and Eq.~(\ref{eq:pta2}), as 
 \bea 
 -\Delta^{4} \delta(t-t_{c}) &=& {d \over dt}\left( -3M_{\rm pl}^{2}H^{2} + {\epsilon \over 2} \dot{\phi}^{2} + \lambda^{3} \phi - 3{\epsilon^{2}  \over \lambda^{3}} H \dot{\phi}^{3}\right) \label{eq:pta3}\\ 
-\lambda^{3} - 3\epsilon H \dot{\phi} + {9 \epsilon^{2} \over \lambda^{3}} H^{2} \dot{\phi}^{2} &=& {d \over dt} \left( \epsilon \dot{\phi}  - {3\epsilon^{2} H \dot{\phi}^{2} \over \lambda^{3}} \right) \,. \label{eq:pta4} 
\eea  

We perform a `pill-box' integration over the domain $t=[t_{c}-\delta,t_{c}+\delta]$ and take the limit $\delta \to 0$ which results in a pair of conditions that $\dot{\phi}$ and $H$ must satisfy at the discontinuity. 
These are 
\begin{eqnarray}  
\label{eq:pta5}
 -\Delta^{4} &=& \left[ -3M_{\rm pl}^{2}H^{2} + {\epsilon \over 2} \dot{\phi}^{2} + \lambda^{3} \phi - 3{\epsilon^{2}  \over \lambda^{3}} H \dot{\phi}^{3}\right]^{t_{c}+\delta}_{t_{c}-\delta} \\
 \label{eq:pta6} 0 &=& \left[ \epsilon \dot{\phi}  - {3\epsilon^{2} H \dot{\phi}^{2} \over \lambda^{3}} \right]^{t_{c}+\delta}_{t_{c}-\delta} \,. 
 \end{eqnarray} 
From Eq.~($\ref{eq:pta6}$) we can infer that $H|_{t_{c} + \delta} \neq 0$. If $H=0$ after the transition, then $\dot{\phi}$ would be continuous at the boundary, and Eq.~($\ref{eq:pta5}$) could not be satisfied. This implies that the metric cannot be identically Minkowski space after a phase transition. Although the spacetime inevitably becomes dynamical in the presence of a time dependent $\Lambda$, for practical purposes the system can still approach Minkowski space as an attractor solution. 
 
We can use the fact that $\phi$ is continuous at $t=t_{c}$, and $H=0$ for $t = t_{c}-\delta$, to simplify the matching conditions: 
 \begin{eqnarray} 
 \label{eq:f1} \epsilon \dot{\phi}_{-} &=&  \epsilon \dot{\phi}_{+}  - {3\epsilon^{2} H_{+} \dot{\phi}_{+}^{2} \over \lambda^{3}}  \\ 
 \label{eq:f2} {\epsilon \over 2} \dot{\phi}_{-}^{2} - \Delta^{4} &=& -3M_{\rm pl}^{2}H^{2}_{+} + {\epsilon \over 2} \dot{\phi}^{2}_{+} - 3{\epsilon^{2}  \over \lambda^{3}} H_{+} \dot{\phi}^{3}_{+} \,,  
 \end{eqnarray} 
where $\dot{\phi}_{-} =  -\lambda^{3}t_{c}/\epsilon + c_{1}$ and $+/-$ subscripts denote the values at $t = t_{c} + \delta$ and $t=t_{c} -\delta$ respectively. 

These equations provide initial conditions for $H$, $\dot{\phi}$ that must be satisfied at $t=t_{c}$. In addition, we require 
the aforementioned field continuity $\phi_{+} = \phi_{-} =  -\lambda^{3}t_{c}^{2}/(2\epsilon) + c_{1} t_{c} + c_{0}$. Then for $t > t_{c}$ the dynamical system is evolved according to 
 \bea 
3M_{\rm pl}^{2} H^2 &=& M_{\Lambda}^{4}+\Delta^{4} + {\epsilon \over 2} \dot{\phi}^{2} + \lambda^{3} \phi - 3{\epsilon^{2}  \over \lambda^{3}} H \dot{\phi}^{3}
\label{eq:pta8}\\ 
-2M_{\rm pl}^{2}\dot H &=& 
{\epsilon^{2} \dot{\phi}^{2} \over \lambda^{3}} \ddot\phi - 3 {\epsilon^{2} \over \lambda^{3}}H\dot{\phi}^{3} + \epsilon \dot{\phi}^{2} \label{eq:pta9}\\ 
0&=&\left( \epsilon  - {6\epsilon^{2} H \dot{\phi} \over \lambda^{3}} \right) \ddot\phi
+3\epsilon H\dot\phi + \lambda^{3} - {3 \epsilon^{2} \over \lambda^{3}} \dot{\phi}^{2} (\dot H+3H^2)\,. \label{eq:pta10} 
\eea

 \begin{figure}
  \centering 
  \includegraphics[width=0.6\textwidth]{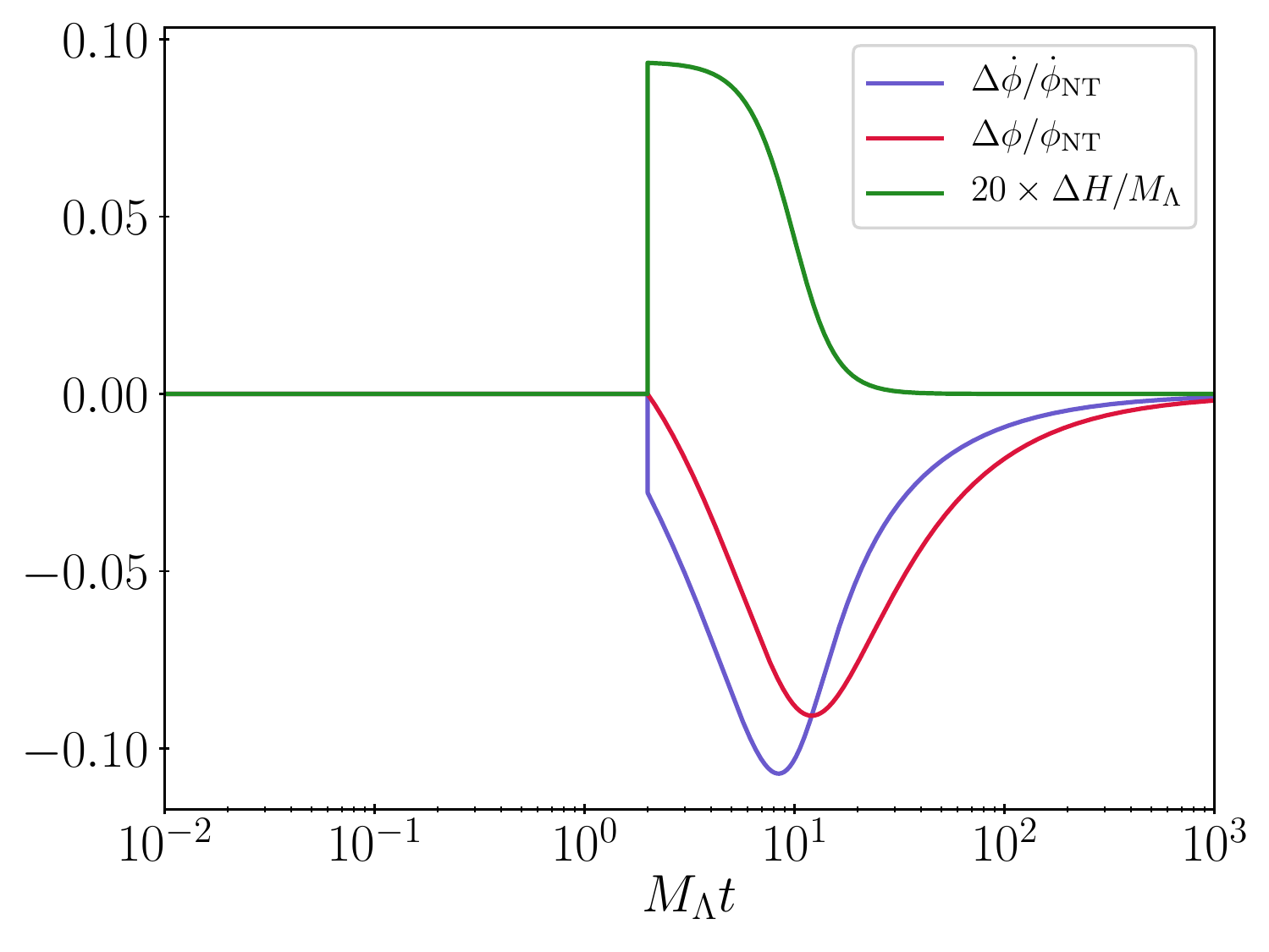}  
  \caption{ 
Despite a vacuum energy phase transition 
$\Delta\Lambda= 0.66 M^4_\Lambda$ 
at $M_\Lambda t=2$, the field dynamically restores the expansion to the Minkowski state $H=0$. The 
fractional differences in $H$ (green), the field evolution $\dot{\phi}$ (blue), and $\phi$ (red), relative to the no transition case, are 
shown vs time. 
  }
  \label{fig:5}
\end{figure}

Figure~\ref{fig:5} shows the transition and the restoration to the Minkowski state by the well tempering. 
To clearly present the effect of the phase transition, we evolve the dynamical system twice with identical initial conditions, the second time with no phase transition by fixing $\Delta = 0$. We then plot the fractional differences $\Delta \phi /\phi_{\rm NT} = (\phi - \phi_{\rm NT})/\phi_{\rm NT}$, $\Delta \dot{\phi} /\dot{\phi}_{\rm NT} = (\dot{\phi} - \dot{\phi}_{\rm NT})/\dot{\phi}_{\rm NT}$, and 
$\Delta H/M_{\Lambda} = (H - H_{\rm NT})/M_{\Lambda}$, where ${\rm NT}$ subscripts denote the `no transition' solution with $\Delta = 0$. We multiply $\Delta H/M_{\Lambda}$ by a factor of $20$ to show its behaviour more clearly. 
 
The dynamical nature of the vacuum energy cancellation mechanism, and the fact that the flat spacetime metric is an attractor solution, combine to preserve the well tempering even through a phase transition where the cosmological constant undergoes an abrupt change in magnitude.

\section{Conclusions} \label{sec:concl} 

Well tempering provides a simple, well defined field theoretic method 
for dynamically removing cosmological constant vacuum energy 
that is generically expected to exist. The approach leaves other, time dependent background energy densities unaffected. Here we examine the case where 
there are no other components, and investigate whether we 
could turn the high energy de Sitter (or anti de 
Sitter) spacetime into Minkowski spacetime, i.e.\ is it actually 
possible to have a flat spacetime? 

Despite the presence of a  large cosmological constant, 
well tempering can make the universe be flat, through a fugue 
of Horndeski functions. This is accomplished without specifying the value of any 
physical constant or the vacuum expectation of any field -- 
it occurs dynamically. The only condition is that the initial 
field conditions on the first time slice, $\phi_{i}$ and  $\dot{\phi}_{i}$, solve the Hamiltonian constraint. It is subsequently solved at all times.

We explore the solutions from combinations of various terms 
from the Horndeski Lagrangian, $K$, $G_3$, and $G_4$ 
(B$^{\flat}$ minor for $K+G_3$ and B$^{\flat}$ major for 
$K+G_3+G_4$ with a coupling to the spacetime curvature), 
specializing to the shift symmetric case for additional 
quantum robustness. After deriving analytic relations and 
giving several examples of 
successful well tempering, we numerically solve the equations of 
motion and exhibit the evolution of the field and approach of 
the cosmic expansion to the zero Minkowski value. 

We present further relations necessary to ensure the lack of  
ghosts and Laplace instability in the theory. 
The latter can be restrictive for large field excursions. 
Finally we 
demonstrate that the dynamics adjusts to a phase 
transition in the vacuum energy, providing a full cancellation both prior to and in the asymptotic future of the transition, 
and preserving Minkowski spacetime.

\acknowledgments

SAA is supported by an appointment to the JRG Program at the APCTP through the Science and Technology Promotion Fund and Lottery Fund of the Korean Government, and was also supported by the Korean Local Governments in Gyeongsangbuk-do Province and Pohang City. 
EL is supported in part by the Energetic Cosmos Laboratory and by the U.S. Department of Energy, Office of Science, Office of High Energy Physics, under Award DE-SC-0007867 and contract no. DE-AC02-05CH11231.


\end{document}